\documentstyle[twocolumn,aps,epsf,prb]{revtex}

\begin{document}
\draft
\title{Teaching Newton with anticipation...}
\author{P. Fraundorf}
\address{Department of Physics \& Astronomy \\
University of Missouri-StL, \\
St. Louis MO 63121}
\date{\today }
\maketitle

\begin{abstract}
Care making only clock-specific assertions about elapsed-time, and 
other ``space-time smart'' strategies from the perspective of a selected 
inertial map-frame, open doors to an understanding of anyspeed motion 
via application of the metric equation. {\bf Keywords:}  
space-time, acceleration, elapsed-time, lightspeed, kinematics.
\end{abstract}



\section{Forward-Looking Introductions}

Introductory texts tell students to specify a coordinate system whenever
displacements are discussed. By asking that the clock's frame of motion be
specified {\em whenever} time intervals are discussed, a habit is begun which 
will be needed to understand motion at high speeds.

For example, if we define a ``map'' frame in which both yardsticks and
clocks reside, then velocity and acceleration defined using only map
distances and times retain their classical form. In terms of these
``coordinate'' quantities, the classical equations of translational motion
listed in Figure 1 make well-defined predictions in space-time.  In fact,
all except four of these predictions are exact not only for low-speed
motion, but for motion at any speed as well!

There is a caveat. Since these equations only concern distances and times
measured from the vantage point of the map, they say {\em nothing} about the
experiences of the accelerated object or traveler. More on this later. The
utility of Fig. 1 for motion at any speed thus assumes that one avoids making 
the assumption, hidden or otherwise, that time passes similarly for everyones'
clocks.  Time elapsed is clock-dependent, save to first order at low speeds.

It is also worthwhile to look at {\em other} implicit assumptions of 
classical mechanics. Figure 2 lists many of the variables familiar from
Newtonian physics, and asks the question: Which of these quantities, when
used to describe motion between two sneezes of an airplane pilot (for
example), did mechanics classically assume are independent of one's choice of
reference (or map) frame?

The answer is that distance between sneezes depends on map-frame (e.g. the
pilot thinks both sneezes occur at the same place, namely the airplanes'
cabin), as does velocity (since a map-frame moving with the pilot would see
the pilot standing still). Momentum and energy obviously depend on velocity,
and hence on choice of map-frame as well.

However, classical mechanics {\em often implicitly} assumes that elapsed times,
observed accelerations, and applied forces are the same for all map-frames.
Some of the teachers polled at two recent AAPT meetings also guessed that
the rate at which energy increases is frame-independent classically.  Rate of 
energy change of course depends on frame, since it equals force times velocity. 
So let's be explicit: For studies involving motion at speeds much less than the
speed of light, of the quantities listed in Fig. 2 one can safely assume 
that only time-elapsed, observed acceleration, and applied force depend very little 
on one's choice of map-frame. For motion studies involving speeds approaching 
lightspeed, {\em all} of the quantities in Fig. 2 depend strongly on 
ones' choice of frame.

Forward-looking habits in the way introductory physics students think about
time, as well as about how quantities will look in one frame or another, are
important in at least two ways. First, they will minimize the things that
students going on in physics will have to {\em unlearn}, like the use of
time in a clock-independent manner. These things are crucial to a solid
understanding of both special relativity and curved space-time. Second, as
described below, such habits open the door to a deeper and simpler
understanding of space-time for students with only one course in
physics. This approach and philosophy is consistent with the proposal by
Edwin Taylor in his 1998 AAPT Oerstead Medal talk, which describes a way to
provide deeper understanding with fewer math prerequisites in a second
physics course. Empowering students in a {\em first course} with deeper
physical intuition also provides us with a significantly better-informed
taxpaying and consuming public.

\section{Equations Good at Any Speed}

Thus students can be taught to be wary of assumptions about frame
invariance when talking about both times and distances, even as unidirectional
motion is introduced via the usual classical expressions. Not
presuming to understand how traveler clocks behave at high speeds, they may
then be eager to learn more about velocity and acceleration at any
speed, well before they are ready for ``multi-frame'' relativity. This may be done
by introducing the metric equation as a space-time extension of Pythagoras' 
Theorem, written so that traveler-time is the invariant. This tells students 
most everything they need to know about traveler time at high speeds.  
However, more sophisticated use of this metric equation is needed before 
students are prepared to predict measurements with traveler yardsticks 
at high speed as well.  Of course, saying ``don't go there'' could pique their 
interest in further studies of relativity and space-time geometry downstream.

Figure 3 introduces proper (or ``traveler'') time with the metric equation,
followed by three corollary variables (gamma, proper velocity, and proper or
``felt'' acceleration). It then provides a set of equations, patterned after
Fig. 1, which are {\em exact} for unidirectional motion at any speed. 
There is a new integral of constant proper acceleration (the
so-called rapidity integral for proper time), and proper force $F_o$
(defined as mass times proper acceleration) has been listed separately from
frame-variant force $F$, since for non-unidirectional motion the two differ.

Lastly, Figure 4 now considers frame-invariance at any speed for the new
variables as well as the variables discussed classically in Figure 2. Note
that the new variables, which arose naturally from the relation for
traveler-time provided by the metric equation, include three true
frame-invariants: proper time, proper acceleration, and proper force. In a
sense, then, the variables whose frame-independence was lost in the first
part of this paper have been reborn in truly frame-invariant form, thanks to
Minkowski's space-time extension of Pythagoras' theorem.

\section{Discussion}

For students who are pictorially-oriented (or equation-shy), a nomogram
plotting all variables versus distance traveled from rest can be put
together with these equations. To illustrate, Figure 5 allows graphical
solution of constant acceleration problems with almost any combination of
input variables in range of the plot. As shown in the caption to Fig. 6, 
for {\em analytical} solution of problems the kinematic equations can be 
compacted into two simple equality strings. 
Other problems with solutions, on-line solvers, a discover-it-yourself maze, 
and a companion derivation of the ``felt'' acceleration equation in Fig. 3, 
may be found on web pages linked to: http://www.umsl.edu/~fraundor/anyspeed.html.

\acknowledgments

This work has benefited indirectly from support by the U.S. Department of
Energy, the Missouri Research Board, as well as Monsanto and MEMC Electronic
Materials Companies. It has benefited most, however, from the interest and
support of students at UM-St. Louis.

\eject

\begin{figure}[tbp]
\caption{Classical equations written with variables defined unambiguously {\em at 
any speed}.  Expressions accurate only at speeds much less than lightspeed are 
outlined.  In these figures the differential operator ``d'' is used like a ``$\Delta $'' 
(which means final minus initial), applied in particular to tiny increments.}
\label{Fig1}
\end{figure}

\begin{figure}[tbp]
\caption{Quantities familiar in the classical study of motion.  
Outlined are quantities classically assumed (often implicitly) to be 
independent of one's choice of map frame.  For motion at sufficiently 
high speeds, however, none of the quantities listed here is at all 
independent of one's frame of motion!}
\label{Fig2}
\end{figure}

\begin{figure}[tbp]
\caption{Equations analogous to the classical equations, which {\em for 
unidirectional motion} are exact at any speed.  The ``non-defining 
equalities'' complicate if motion is polydirectional.  Traveler-time is 
introduced using the metric equation%
\protect\cite{Minkowski} as a space-time extension of Pythagoras' theorem
while, in the spirit of J. Bell\protect\cite{Bell} and classic pedagogical
tradition sticking to one reference frame with its tangible Cartesian map
and extended array of synchronized clocks. Frame-invariant proper 
acceleration\protect\cite{TaylorWheeler} is very simply expressable for 
unidirectional motion in context of a single map-frame\protect\cite{AnySpeed}. 
{\bf Note:}  Comparing the rates of clocks (e.g. when defining $\gamma$ as a ``speed 
of map-time'') requires that simultaneity, a property of the relationship between 
events that is strongly {\em frame-dependent} at high speeds, have specified 
meaning.  Choice of a map-frame provides an unambiguous definition of 
simultaneity, namely that for map-frame observers.  However, the relative 
clock rates experienced by observers moving with respect to the map (like an 
accelerated traveler) will differ.  
}
\label{Fig3}
\end{figure}

\begin{figure}[tbp]
\caption{Quantities useful in the study of motion at any speed, including 
three outlined variables whose value is {\em truly independent} of one's choice of 
map-frame.  For problems at high speed on a human scale, convenient units for 
time are years [y] and traveler years [ty], while convenient units for distance 
are lightyears [ly].  In this case coordinate-velocity has units of {\em %
lightyears per map-year} or``c'', while proper-velocity\protect\cite
{SearsBrehme} (which has no upper limit\protect\cite{Shurcliff}) is in {\em %
lightyears per traveler-year}. One lightyear per traveler-year marks
the transition to high speed nicely (it corresponds to $v \simeq 0.707 c$),
but it is so cumbersome to say that a mnemonic may 
help. For example, our ``land-speed record'' for $%
m > 0$ objects is held by CERN electrons traveling at $\simeq 10^5[ly/ty]$!  
Another advantage of these units of course is that, by happy coincidence, 
an acceleration of 1 ``gee'' (something we know humans can live with) is 
about $1 [ly/y^2]$.}
\label{Fig4}
\end{figure}

\begin{figure}[tbp]
\caption{An ``anyspeed'' constant acceleration nomogram. If distance is in
lightyears and time in years, then 1 ``gee'' proper-accelerations correspond
to $\alpha = 9.8[m/s^{2}] \simeq 1.03[ly/y^{2}] \simeq 1[ly/y^{2}]$. Thus
plotting a line up from the number 2 on the horizontal axis above allows one
to determine, by inspection, the following results of a 1 ``gee'' trip from
rest over a distance of $2[ly]$: final coordinate velocity ($\simeq
0.9[ly/y]$), elapsed traveler time ($\simeq 1.7[ty]$), elapsed map time ($%
\simeq 2.8[y]$), and final proper velocity ($\simeq 2.8[ly/ty]$). The
``chase-plane'' parabola follows Galileo's low-speed curve (cf. Fig. 1).}
\label{Fig5}
\end{figure}

\begin{figure}[tbp]
\caption{Anyspeed kinematics for unidirectional motion are easily summarized 
by two equation strings. These are the ``velocity conversions'' {\bf I}: $\gamma
=\frac 1{\sqrtsign{1{\text -}(v/c)^2}}=\sqrtsign{1+(w/c)^2}=\cosh [\eta ]$ and the ``motion
integrals'' {\bf II}: $\alpha =\frac{\Delta w}{\Delta t}=c\frac{\Delta \eta }{%
\Delta \tau }=c^2\frac{\Delta \gamma }{\Delta x}$. For the first of four
legs of this trip, \emph{proper-acceleration} $\alpha =1$``gee''$\simeq 1.03$%
[ly/y$^2$] and traveler-time elapsed $\Delta \tau =\frac{56.46}4=14.115$%
[ty]. Since speed starts at zero, from (II) rapidity $\eta _{final}=\Delta
\eta =\frac{\alpha \Delta \tau }c\simeq 14.538$. From (I), proper-velocity 
$w_{final}= c \sqrtsign{\cosh [\eta _{final}]^{2}-1}= c\sinh [\eta _{final}]\simeq
1.03\times 10^6$[ly/ty], and from (II) map-time elapsed $\Delta t=\frac{%
w_{final}}\alpha =\frac c\alpha \sinh [\frac{\alpha \Delta \tau }c]\simeq
10^6$[y]. Combining all 4 legs gives total earth-time elapsed on return of 
$4\times 10^6$ years, even though the traveler is only 56.46 years longer in
the tooth!}
\label{Fig6}
\end{figure}

\end{document}